\documentstyle[11pt,aaspp4]{article}
\def\arcs{\char'175\ ~}
\def\arcsc{\char'175 }

\def\hub{\ifmmode H_\circ\else H$_\circ$\fi}
\def\kms{~km~s$^{-1}$\ }

\def\dug{$\rm ^o~$}
\def\dugc{$\rm ^o$}
\def\etal{et~al.\ }

\journalid{337}{15 January 1989}
\articleid{11}{14}
\begin{document}

\title{On the Origins of Starburst and Post-Starburst Galaxies in Nearby 
Clusters\altaffilmark{1}}

\author{Nelson Caldwell}
\affil{F.L. Whipple Observatory, Smithsonian Institution, Box 97, Amado AZ 
85645}
\affil{Electronic mail: caldwell@flwo99.sao.arizona.edu}

\author{James A. Rose}
\affil{Department of Physics and Astronomy, University of North Carolina, Chapel 
Hill, NC 27599}
\affil{Electronic mail: jim@physics.unc.edu}

\author{Kristi Dendy}
\affil{Department of Physics and Astronomy, University of North Carolina, Chapel
 Hill, NC 27599}
\affil{Electronic mail: dendy@physics.unc.edu}


\altaffiltext{1}{Based on observations with the NASA/ESA {\it Hubble Space
Telescope} obtained at the Space Telescope Science Institute, which is
operated by AURA, Inc., under NASA contract NAS 5-26555.}

\begin{abstract}
{\it HST} WFPC2 images in B (F450W) and I (F814W)
have been obtained for
three starburst (SB) and two post-starburst (PSB) galaxies in the Coma cluster,
and for three such galaxies in the cluster DC2048-52.
V (F555W) and I images for 
an additional PSB galaxy in Coma have been extracted from the {\it HST} archive.
Seven of these galaxies were previously
classified as E/S0 on the basis of ground-based images, one as Sa, and the
other as an irregular. 

The {\it HST} images reveal these SB/PSB galaxies to be heterogeneous in 
morphology.  Nevertheless a
common theme is that many of them, especially the SB galaxies, tend to have
centralized spiral structure that appears simply as a 
bright ``bulge''on ground-based images.  In addition, while some PSB galaxies
exhibit distinct spiral
structure, on the whole they have smoother morphologies than the SB
galaxies.  The morphologies and luminosity profiles are generally consistent
with substantial starbursts in the form of centralized spiral structure (the SB
galaxies) which fade into smoother morphologies (the PSB galaxies), with 
lingering spectroscopic evidence for past central starbursts.  An
important point is that the PSB galaxies retain disks, i.e, they have not 
evolved into spheroidal systems.

While the morphologies revealed in the {\it HST} images are heterogeneous, and thus
may not fit well into a single picture, we see evidence in several cases that
the morphologies and centralized star formation have been driven by external
tidal perturbations.   We discuss several physical mechanisms for inducing
star formation in cluster galaxies with a view towards explaining the
particular morphologies seen in the {\it HST} images.

\end{abstract}

\keywords{galaxies: clusters --- galaxies: clusters (Coma,DC2048-52) ---
galaxies: evolution --- galaxies: interactions --- galaxies: starburst}

\section{Introduction}

Numerous studies have convincingly established that at z$>$0.3 rich clusters 
of galaxies have
a higher fraction of photometrically blue galaxies than is found in present-epoch
clusters of similar richness (e.g., Butcher \& Oemler 1978, 1984; Couch \&
Newell 1984; MacClaren \etal 1988; Dressler \& Gunn 1992; Rakos \& Schombert 
1995; Belloni \etal 1995).  In addition, follow-up
spectroscopy (e.g., Dressler \& Gunn 1982, 1983, 1992; Lavery \& Henry 1986;
Henry \& Lavery 1987; Couch \& Sharples 1987; Soucail \etal 1988; Pickles \&
van der Kruit 1991) has revealed that many galaxies in distant clusters exhibit 
strong emission lines, indicating either ongoing star formation or
starburst (SB) activity, while others
show strong Balmer absorption lines with little or no emission, 
indicating either recently truncated star formation or
poststarburst (PSB) activity.  The fraction of the strong Balmer line galaxies which
have experienced a major starburst as opposed to sudden truncation of star
formation is a matter of current debate (cf. Barger \etal 1996; Balogh \etal 
1997; Morris \etal 1998).
Recently, morphological studies carried out with the {\it Hubble Space 
Telescope} have shown that a larger fraction of the star forming 
galaxies are classified as
spiral, irregular, or interactions/mergers than in nearby clusters (Dressler 
\etal 1994a,1994b; Couch \etal 1994; Wirth \etal 1994; Couch \etal 1998),
a result that confirms the ground-based imaging studies of Thompson
(1988), Lavery \& Henry (1988), and Lavery \etal (1992).  

As mentioned above, at z=0 there are relatively few blue galaxies in rich
clusters, i.e., with B-V$>$0.2 mags
bluer than the c-m sequence of the red galaxies in the cluster 
(e.g., Bower \etal 1992).  Furthermore, in present-epoch clusters
only a small fraction of galaxies are classified as spirals, at least within the
morphological framework that has been developed from mostly isolated nearby 
spirals.  In 
fact, even those cluster galaxies classified as spirals are usually found to be 
depleted in HI content (e.g., Haynes \etal 1984 and references therein;
Giovanelli \& Haynes 1985; Gavazzi 1989).  Nevertheless, recent work has
indicated that the differences in evolutionary processes between nearby and
distant rich clusters may not be as extreme as previously considered.  Moss \&
Whittle (1993) and Moss \etal (1995) found that a higher fraction of spirals in
nearby clusters that are morphologically classified as early type spirals show
elevated, and centrally concentrated, star formation levels than in a field
sample.  Gavazzi \etal (1995) also found evidence for elevated star
formation rates in cluster spirals.  In addition, Caldwell et al (1993, 1996) 
and Caldwell \& Rose (1997,
1998) have shown that a substantial fraction of early-type galaxies (i.e.,
those morphologically classified as E or S0) in nearby rich clusters
show signs of PSB or SB
activity.  Finally, Koopmann \& Kenney (1998) have pointed out that in
general spiral galaxies in clusters do not fit consistently within the 
morphological framework of isolated spirals.  Specifically, many cluster spirals
with low disk star formation rates that would indicate 
early-type spiral morphologies in fact have concentrations of light typical of
late type spirals.  Thus there may exist many late
type spirals in clusters at low redshift which have such suppressed star 
formation rates
in their disks (due to the gas depletion processes present in rich clusters),
that they are lumped together with early type spirals.  

In short, a perhaps more
accurate picture of nearby clusters is that environmentally
driven evolutionary processes in galaxies 
are still ongoing.  These environmental processes may be
qualitatively similar to those observed in distant clusters, although they
are occurring at a reduced level in intensity and 
frequency. The reduced intensity 
in nearby clusters
may be due in part to the depletion in HI reservoir in the general galaxy population
since z=0.3, due to star formation or other removal processes. 
More generally, the reduction in intensity and frequency has been discussed
within the context of a hierarchical clustering scenario for structure formation
by Bower (1991) and Kauffmann (1995).

Given that evolutionary processes in cluster galaxies, perhaps similar in nature to those
occurring in distant clusters, are indeed ongoing at the present epoch, then
the proximity of nearby rich clusters, such as Coma, provides an excellent 
prospect for studying these processes in detail.  In particular, the star
formation bursts can be studied at high enough spatial resolution that their
cause(s) might be revealed.  For example, how centrally concentrated are the
starbursts?  Are there any young star clusters, such as are found in merging
galaxies (e.g., NGC 1275,  Holtzmann \etal 1992; NGC 7252, Miller \etal 1997).
Are there morphological peculiarities associated with the
starbursts that can be unambiguously linked to tidal disturbances or other
external triggering mechanisms?  
And what are the morphological differences
between SB and PSB galaxies? To that end, we have otained WFPC2
images with the {\it Hubble Space Telescope} ({\it HST})
of a number of SB and PSB galaxies in the Coma and
DC2048-52 nearby rich clusters.  The morphological characteristics of these
galaxies, as revealed by WFPC2 imaging, forms the subject of this paper.  The
characteristics of the {\it HST} images are described in \S 2.  In \S 3 the
morphologies and inner luminosity and color profiles of the Coma and 
DC2048-52 SB/PSB galaxies are presented.  In \S 4 we discuss 
which mechanism can best account for the SB/PSB
activity and morphological disturbances revealed by the {\it HST} images.

\section{Observational Data}

Before describing the {\it HST} WFPC2 images, we briefly summarize how our
sample of nine galaxies was selected.  
We originally obtained multi-fiber spectroscopy for many galaxies in Coma and
DC2048-52 (and in three other clusters) which we believed to be early-type
based on morphologies from ground-based images and/or colors (Caldwell \etal 
1993; Caldwell \& Rose 1997).  For a substantial
fraction (typically $\sim$15\%) of these galaxies the spectra indicate
either current or recent star formation.  The nine galaxies chosen 
were then selected from among those with spectroscopic evidence for unusual 
star formation histories.  Furthermore, they were chosen to cover a 
representative range of current and past star formation.  Finally, while
in Coma the SB/PSB galaxies are primarily located in the SW region of the
cluster, in the case of DC2048-52 they are much more uniformly spread over
the main cluster and the northern subcluster.

WFPC2 images were obtained in the B (F450W filter) and I (F814W filter) 
passbands with {\it HST} during Cycle 6 for four fields in the Coma cluster
and three fields in DC2048-52.  In each case a SB or PSB galaxy was
centered in the PC frame, except in the case of one field in Coma, where
both the PSB galaxy Dressler 99 and the SB galaxy Dressler 100 were included 
on a single
PC frame.  Two 600 second exposures were 
obtained for each galaxy in the F450W filter, and two 400 second
exposures in the F814W filter.  The images were processed using the 
standard {\it HST} pipeline routines.  To remove cosmic rays, the pairs of
exposures were coadded using the `gcombine' routine in IRAF/STSDAS with the
`crreject' option.  In addition, images of the PSB galaxy Dressler 216 in Coma
were extracted from the {\it HST} archive.  In this case, the galaxy is located
on one of the WF chips, and the passbands are V (F555W) and I (F814).  There
are seven 900 second exposures in V and eight 900 second exposures in I.  The 
same processing steps were used for these images.

Radial light profiles in both colors were extracted for these galaxies using a
program that fits elliptical isophotes (Caldwell \& Rose 1997).  The elliptical
isophotes were found on the I filter image; intensity data falling on these
ellipses were then collected from the bluer image (B or V), after adjusting for
a change
in center.  Calibrations for the F450W, F555W, and F814W frames were those 
provided in Holtzmann \etal (1995).
Error estimates in the surface brightnesses, magnitudes, and colors
were
calculated from the photon statistics, the readout noise, and the error in
determining
the mean sky.  Color profiles were found by subtracting the surface brightness
profiles
of the two filters.  No attempt at matching the point-spread functions of the
two
filters was made.  The mean color quoted in Table \ref{tab1} refers to the integrated
color
within the radius where the I surface brightness falls to 23 mag arcsec$^{-2}$.
For reference,  normal luminous elliptical galaxies have B-I$\sim$2.3.

In addition to the {\it HST} images, ground-based images for eight of the
nine galaxies
have been obtained, either with the 1.2-m telescope of the F. L. Whipple 
Observatory for the Coma galaxies or with the CTIO 0.9-m telescope for the
DC2048-52 galaxies.  The characteristics of these ground-based images are
given in Caldwell \etal (1993) and Caldwell \& Rose (1997).

The {\it HST} and ground-based images are displayed in Figs. \ref{pic2},
\ref{pic1}, and \ref{pic3}.  In Fig. \ref{pic2} we have assembled those galaxies
with clear signs of active star formation, while galaxies with
relatively smooth morphologies, i.e., little or no sign of ongoing
star formation, are shown in Fig. \ref{pic1}.  Galaxies which appear to be 
weak late-type
spirals are shown in Fig. \ref{pic3}.  Each row in these Figures shows the
{\it HST} images at low and high contrast in the first two panels, followed
by the ground-based image in the third panel.

As mentioned earlier, the SB or PSB nature of the nine galaxies was originally 
determined from multi-fiber spectroscopy with either the Hydra spectrograph on
the KPNO 4-m (for Coma) or the Argus spectrograph on the
CTIO 4-m (for DC2048-52).  The one exception is Dressler 61 in Coma, for which the
PSB nature was determined from a spectrum obtained with the MMT.  All of these
observations are described in Caldwell \etal (1993) and Caldwell \& Rose (1997).
In addition, the SB or PSB nature was shown to be centrally concentrated,
although spatially extended, from follow-up long-slit
spectroscopy of Coma galaxies described
in Caldwell \& Rose (1996).  For D61 in Coma, an additional spectrum was 
obtained with the 1.5-m telescope of the F. L. Whipple Observatory and FAST 
spectograph (Fabricant \etal 1998) at 1.5 \AA/pixel and covering the wavelength 
region $\lambda$$\lambda$3635-7585 \AA.  These spectroscopic observations are 
referred to in \S 3 below, where the properties of the nine SB/PSB galaxies are
summarized.

\section{Morphologies and Luminosity Profiles of Starburst and Poststarburst 
Galaxies}

The morphologies and luminosity profiles of the SB and PSB galaxies for which we
have HST images are quite diverse.  As a result, we begin with an
individual discussion of each galaxy, and in \S 4 draw general conclusions
about the starburst nature of these galaxies.  For convenience of organization,
we begin with the SB galaxies of Fig. \ref{pic2}, then proceed to the PSB
galaxies of Fig. \ref{pic1}, and finish with the weak late-type spirals of
Fig. \ref{pic3}.  The basic parameters for each galaxy discussed below are
summarized in Table \ref{tab1}.
For the purposes of the following discussion, it may be useful to know that
in Coma, which we assume to be 100 Mpc distant (corresponding to
\hub = 70 \kms Mpc$^{-1}$, and farther than the 86 Mpc we assumed in
previous papers),  1\arcs= 490 pc, and for DC2048-52 (at 200 Mpc), 
1\arcs= 980 pc.  Hence the multi-fiber spectra, which cover 3\arcs in diameter
on the Coma cluster galaxies and 2\arcs diameter on the galaxies in DC2048-52,
provide integrated spectra of typically the central $\sim$2-3 kpc in diameter 
of these galaxies.

\subsection{Starburst Galaxies}

\noindent {\it D15 in Coma:}\\
The spectrum of Dressler 15 in Coma (hereafter referred to as Coma-D15) 
reported in
Caldwell \etal (1993) and Caldwell \etal (1996) reveals strong emission lines
that are a mixture of an HII region spectrum and an AGN spectrum.  The
emission is very centrally concentrated, but spatially resolved out to
about 2\arcsc, and has a velocity 
gradient with a full amplitude of 100 \kms.  Thus
Coma-D15 is currently undergoing a central starburst and also harbors an AGN.
A nearby apparent ``companion'' galaxy is in fact a 
background galaxy with redshift of z=0.127.

On the basis of ground-based images, Coma-D15 has been classified as an S0
galaxy (Dressler 1980, Caldwell et al. 1993, Andreon et al. 1996).
The WFPC2 images in Fig. \ref{pic2}, however, reveal that the central bright
``bulge''
is resolved into a
spiral disk, where the
distance from the nucleus to the inner bright spiral arms is $\sim$1.5\arcsc,
or $\sim$0.75 kpc. The spirality extends to only 3\arcs (1.5 kpc), which 
explains why it was missed on the ground-based images.
This bright inner
spiral structure is similar to that seen with {\it HST} of some other S0 
galaxies, e.g. NGC 524 and NGC 3599
(Lauer \etal 1995), though the spiral in Coma-D15 is stronger, and
better resembles that seen in the merger product galaxy NGC 7252
(Whitmore et al. 1993).  In fact,
the radial extent of the spiral arms in both galaxies is very similar
($\sim 1.5$ kpc).  There are however no large tidal tails in Coma-D15,
as are seen in NGC 7252, nor is there a companion galaxy evident.

The radial luminosity profile
of Coma-D15 is shown in Fig. \ref{sb_pro}. where the luminosity profiles
in B and I are plotted versus radius ($R$) and $Log(R)$.  The B-I color
profile is also plotted.  The thin solid lines represent
$\pm$1 $\sigma$ errors for the B light and color profiles.
An additional systematic error from the calibration may be
present at the level of 5\%, which has no effect on our results.  The radial
luminosity profile
shows a strongly peaked light distribution inside the central 1\arcs in
radius, then a bright ``bump'' in the profile caused by the spiral arms, and
then a basically exponential disk beyond a radius of $\sim$2\arcsc.  The
radial color profile reveals a blue nucleus
surrounded by a substantially redder (by $\sim$0.4 mag in B-I) disk, which then
gives way to the blue spiral arms at $\sim$1.5\arcsc.

\noindent {\it D100 in Coma:}\\
Dressler 100 in Coma (hereafter Coma-D100) is an apparent nearby companion to
Coma-D99 which has been classified as an irregular and an SBa from ground-based images
(Dressler 1980, Andreon et al. 1996, respectively).  
Although Coma-D99 and Coma-D100 are projected to be only 17\arcs (8.5 kpc) apart, they
have a large velocity difference of 4700 \kms.  Thus, even if they are indeed
experiencing a close passage, they cannot be a strongly
interacting pair in the traditional sense of low encounter velocity 
interacting/merging galaxies.  The long-slit spectrum of Coma-D99 from
Caldwell \etal (1996) was oriented at a position angle of 121$^{\circ}$ E of N, in
order to pass through the center of Coma-D100 as well.  This spectrum reveals
that Coma-D100 has an active star formation spectrum in the central 
$\sim$2\arcs ($\sim$1 kpc) in radius, i.e., [OIII]$\lambda$5007 and H$\beta$
emission can be seen out to this radius; strong underlying Balmer absorption
can also be seen in
the higher order Balmer lines.  At $\sim$3\arcs radius, the spectrum is
characteristic of a PSB state, i.e., no emission is detected but strong Balmer
absorption lines are clearly evident.  
We have applied the Ca II H + H$\epsilon$ versus H$\delta$/Fe I age dating 
technique that was developed in Leonardi \& Rose (1996) and applied to other
Coma cluster PSB galaxies in Caldwell \etal (1996) and Caldwell \& Rose (1998).
This technique, which decouples the mostly degenerate effects of PSB age from
burst strength (for a fading burst superposed on an underlying ``old''
galaxy), relies on the high sensitivity of the relative strength of H$\epsilon$
versus the Ca II H and K lines to the contribution from A stars.  
Using this age dating method,
we find that the age of the outer PSB region of Coma-D100, summed
over both sides of the slit, is $\sim$0.25 Gyr post starburst (see Fig. \ref{age_ps}).
Thus, while
star formation is still proceeding vigorously in the central 1 kpc of
Coma-D100, it has already subsided at a radius of $\sim$1.5 kpc.  Such
is seen in other nearby SB galaxies such as NGC5253 (Caldwell \& Phillips
1989).
Our WFPC2 images certainly support an unusual morphology for Coma-D100 (see
Fig. \ref{pic2}).  A strong two-armed spiral pattern is visible as well as
highly
visible dust lanes.  A bright nucleus is partially obscured by heavy dust
in the central $\sim$0.5\arcs in radius.  The two-armed spiral pattern is
visible out to 3\arcs (1.5 kpc) in radius.  A set of parallel dust lanes
project out on the east side of Coma-D100, extending out to $\sim$3.4\arcsc.
Dust predominates on the west side in the central $\sim$0.5\arcs in radius, but
is more prominent on the east side at larger radii.  The dust shows up
prominently in
the color profiles, causing a redder color around 1-1.5\arcs than the colors
inward and outward of that location.

Finally, from an MMT long-slit spectrum we measured a velocity difference of
only 14 \kms between the NW and SE side of the slit at a distance of 3\arcs
from the nucleus, which is within the measurement
errors of $\sim$$\pm$20 \kms.  Thus we can conclude that there is little
rotation at the 121$^{\circ}$ position angle. This is smaller than we would
expect to see for a disk galaxy measured  $\sim$63$^{\circ}$ away from the
galaxy's major axis, though we hestitate to call it unusual.

\noindent {\it D45 in Coma:}\\
Dressler 45 in Coma (hereafter Coma-D45) was originally determined to be a
starburst galaxy in Caldwell \etal (1993), because of
its strong emission lines characteristic of an HII region spectrum as well as
enhanced Balmer absorption that is evident in the higher order Balmer lines.
The kinematics of the ionized gas was studied in Caldwell \etal (1996), where
an asymmetric rotation curve was discovered from long-slit MMT spectra at
four position angles.  Coma-D45 was classified as Sa from ground-based images,
and an R band image presented in Caldwell \etal (1996) tended to confirm that
morphology, while revealing an asymmetric structure to the outer isophotes.
We must correct an error in the color reported in Caldwell et al. (1993).
The B-R color of Coma-D45 from the GMP paper was thought to be quite red,
unusual for an Sa galaxy (Figure 12 in Caldwell et al.
1993).  A coding error was the cause of this statement: in fact GMP did not
report a color for Coma-D45. A similar problem occured for Coma-D15; there were
no
other such errors.  The colors for both Coma-D45 and Coma-D15 are bluer than
previously reported, substantially so in the former case, as can be surmised
from
Table \ref{tab1}.

The WFPC2 images of Coma-D45 reveal that the central ``bulge'' of this galaxy
is actually a highly irregular structure, with several bright knots (presumably
HII regions) forming a partial ring with a diameter of $\sim$1\arcsc, or about
0.5 kpc.  As can be seen in Fig. \ref{pic2}, the structure of Coma-D45 is
irregular on all scales, including the fainter isophotes mentioned above in
the ground-based images.  The radial luminosity profile (Fig. \ref{sb_pro})
shows two exponentials of different scale lengths, with the break
occurring at $\sim$2-3\arcsc.  The highly irregular morphology of Coma-D45
revealed in the WFPC2 images may clarify some of the asymmetric nature of the
rotation curve reported in Caldwell \etal (1996), in that the kinematic center
of
the galaxy may be well displaced from the emission line peak intensity.  In
that case, the rotation curve can be seen as reasonably symmetric around a
displaced center within the central $\sim$2\arcs in radius.  Outside of that
radius, however, the gas appears to be counter-rotating.

\subsection{Poststarburst Galaxies}

\noindent {\it D99 in Coma:}\\
High S/N MMT spectroscopy of Dressler 99 in Coma (hereafter Coma-D99) shows 
that it is a PSB
galaxy seen $\sim$1.0 Gyr after the starburst.  The long-slit MMT 
spectra showed
that the strong Balmer absorption lines are confined to the inner 1\arcs in
radius.  On ground-based images the galaxy has been classified as S0 or E/S0.
(Dressler 1980, Caldwell \etal 1993).

The WFPC2 images (see Fig. \ref{pic1} for the poststarburst galaxies) 
show a smooth light distribution, 
with no
indications of spiral structure.  The bright nuclear ``bulge'', which is
$\sim$0.5\arcsc, or $\sim$0.25 kpc, in radius, is surrounded by what appears to
be a smooth lower surface brightness envelope.  The radial luminosity profile
(Fig. \ref{psb_pro})  confirms the bulge/disk character of the galaxy. 
The radial
color profile reveals the central area (but outside of 0.2\arcsec)
to be 0.3-0.5 mag bluer in B-I than the
surrounding disk, thus confirming the earlier conclusions from ground-based
imaging and spectroscopy that the starburst was highly concentrated in the
center of Coma-D99.  The inner few pixels ($\sim $0.2\arcsc) of the B
image show some structure that may indicate the presence of an off-nuclear
star cluster (this feature appears on both B images, and thus is not
an artifact).  Interestingly, the I image shows a more peaked nucleus than 
the B image, which is unresolved in a deconvolved image.  A nuclear star
cluster may be present, somewhat reddened, for the color of the nucleus
alone (within 0.1\arcsc) is B-I=1.6, substantially redder than the 
surrounding area.

Interestingly, the central surface brightness of Coma-D99 is quite low,
at around B=18.3 mag arcsec$^{-2}$, or an estimated V=17.6 mag arcsec$^{-2}$.
The expected value for a moderate luminosity E/S0 like Coma-D99 is much
higher, around V=15 (as is found for an archival sample of E/S0's observed 
with HST).  Only high luminosity galaxies such as NGC 4889 tend to have similar
central surface brightnesses.  The fact that a nucleus shows up only in the
I band may indicate that reddening is partly responsible for the low
observed central B surface brightness, though high reddening is 
not indicated by
the overall blue color of the central regions.  The low observed central 
surface brightness must for the most part be real, and may be a clue to 
the mechanism for the starburst in this galaxy.

\noindent {\it D216 in DC2048-52:}\\
On the basis of ground-based images, Dressler 216 in DC2048 (hereafter
DC2048-D216) is classified as an S0.  It was classified as PSB by Caldwell \&
Rose (1997) on the basis of strong Balmer absorption lines and only a small
and uncertain amount of [OII]$\lambda$3727 emission.  The WFPC2 images reveal
a very smooth and mostly symmetric morphology (see Fig. \ref{pic1}).  The only 
asymmetric structure is a slightly curved barlike (i.e., integral sign) 
structure that starts outside the bright nucleus and can be traced out to
$\sim$1.2\arcs in radius (1.2 kpc).  Beyond 1.2\arcs in radius, the morphology
is smooth with no asymmetry. 
The radial luminosity profile (Fig. \ref{psb_pro}) is that of a disk galaxy
with a bulge within 1\arcsec.
The color profile indicates that the
disk becomes slightly bluer towards the center, while the bulge may be yet
bluer, though the different resolutions in the B and I frames may be affecting
the color profile there.

There is a companion spiral to this galaxy, not shown in the figure, located
about
11\arcs away, but for which we have no redshift.  This galaxy is normal in
appearance, without any signs of tidal interactions, so we may presume it
did not interact with DC2048-D216 and cause the latter's starburst.

\noindent {\it D61 in Coma:}\\
High S/N spectroscopy of Dressler 61 in Coma (hereafter Coma-D61) with the MMT
reveals a very strong PSB spectrum, which is extended in radius out to at
least 3\arcsc, and which has a small (4 \AA \ equivalent width) 
component of
[OII]$\lambda$3727 emission.  From an integrated spectrum obtained with the 
1.5-m telescope at Mt. Hopkins over the central 36\arcs in diameter, and
covering H$\alpha$ and [NII]$\lambda$6548,6584, we found that the
equivalent width of the combined [NII]$\lambda$6548 and $\lambda$6584 
emission is $\sim$7.5 \AA.  H$\alpha$ absorption is seen to be exactly balanced
by emission.  From the equivalent widths of the higher order Balmer absorption
lines (which are relatively free of emission) we estimate that the underlying
H$\alpha$ absorption is $\sim$5 \AA.  Thus the ratio of [NII] to H$\alpha$ 
emission is slightly greater than 1, which is typical of LINER emission line
spectra, and definitely in excess of even the most metal-rich HII regions
(e.g., Baldwin \etal 1981; Veilleux \& Osterbrock 1987).
Thus the weak emission spectrum is due to nuclear activity, and not to a
small rate of residual star formation, which strengthens the case that Coma-D61
is indeed a PSB galaxy. 
Using the previously mentioned Leonardi \& Rose (1996)
age dating technique, we have determined the PSB age of Coma-D61 
to be $\sim$0.5 Gyr,
as can be seen in Fig. \ref{age_ps}.  This is a considerably shorter PSB time
than the typically 1 Gyr for other PSB galaxies in Coma to which we have 
applied this technique.  

From ground-based imaging Coma-D61 has been classified as S0.  The WFPC2 images
(Fig. \ref{pic1})
reveal it as an edge-on disk galaxy with a very bright nucleus.  Two radial
dust lanes are apparent about 1\arcs to the east of the center.
Coma-D61 has an unusual bulge: it is
boxy or peanut-shaped, and additionally has an X morphology that is either
part of the box shape, or superposed on that.
A popular parameter to characterize the degree of boxiness is the A4 parameter,
the cos (4$\theta $) deviation from a perfect ellipse, normalized to
the semimajor axis and local intensity gradient.  A4 is about -0.02 in
the bulge area on Coma-D61.

After deconvolving the image using the Lucy-Richardson technique 
(Lucy 1974; Richardson 1972), it is apparent that the nucleus is unresolved
(see Fig. \ref {d61.nuc}),
with the surface brightness steeply falling from B=12.8 mag arcsec$^{-2}$ to
16.1 within 0.09\arcsc. It is not clear how much the AGN continuum
adds to the central luminosity, but if the light comes purely from 
a nuclear star cluster, its luminosity is M$_{\rm B}=-16.3$ within 
3 pixels (0.14\arcs or 66 pc), which represents about 7\% of the galaxy's
total light.  This luminosity is higher than any of the star clusters formed
in the merging galaxies studied by Whitmore et al. (1997), or by
Carlson et al. (1998), and is also higher than the nuclear star
clusters found in normal early-type galaxies by Lauer et al. (1995).
The color of the nucleus is B-I=1.20, about 0.2 mag
bluer than that of the surrounding bulge.

The X-shape is much more clearly seen in Fig. \ref{d61}, where the overall
axisymmetric component of the bulge light
profile has been subtracted off, by fitting ellipses to the isophotes
(the disk light was excluded from the fit).  The
X-shape can be traced out to at least 1.8\arcs (0.9 kpc) outside the nucleus.
The X is not affected by diffraction spikes from the bright nucleus because 
the opening
angle of the X is 110\dugc, not the 90\dug expected from the orthogonal
secondary support vanes.  Moreover, the X is better aligned with the disk of
Coma-D61 than it is with the known orientation of the diffraction spikes on
WFPC2 images.

The disk is also unusual, in that it does not extend into the center.  This
feature
may be what is referred to as a lens (Kormendy 1992), in that a shelf, or constant
surface brightness area in the profile is present, at least in the unprocessed
profile (Fig. \ref{d61_pro}).  In the image in which the bulge has been removed, the profile
along the major axis does not show a shelf for the disk, rather the area shows
a rise and fall in surface brightness with radius.  The maximum brightness
occurs
at the same radius on both sides of the nucleus, so the feature is most likely
a ring rather than a spiral arm.  This ``ring'' has about the same color as the
bulge at that radius (B-I=1.5). On the other hand, the X feature appears to
be somewhat bluer, with B-I=$1.05 \pm 0.10$, though there is a strong systematic
error in this value because of the difficulty in subtracting off the elliptical
bulge in exactly the same way for the images in the different filters.
Still, the blue color supports the idea that the stars contributing to the
X-like distribution are young.

Finally, the MMT spectrum mentioned above was oriented along the
major axis of Coma-D61, and even though the spectrum was not very deep,
we have extracted both stellar and gas kinematics from it.  The stars rotate
with a velocity of about 65 \kms out to 3\arcs (1.5 kpc) from the nucleus.
Similarly, the gas shows a rotation of about 50 \kms.  It would be interesting
to pursue the kinematics of this object with a more detailed study.

\noindent {\it D216 in Coma:}\\
Multifiber spectroscopy of Dressler 216 in Coma (hereafter Coma-D216) indicates
a weak PSB nature for this galaxy, which has been classified as Sa by Dressler 
(1980) from ground-based images.  
The archival WFPC2 V (see Fig. \ref {pic1}) and I
band images
reveal a large bulge as expected for its classification,
and well-defined smooth spiral structure out to a radius of nearly 14\arcsc
(7 kpc).  The radial luminosity profiles indicate that the disk component
only dominates outside a radius of 4\arcs (see Fig. \ref{psb_pro}).  In
addition, the bulge is
considerably redder than the disk, which appears to be consistent with its
weak PSB status.

A final comment regarding Coma-D216 is that the archival {\it HST} image of the
bright ``E+A'' galaxy in A665 at z=0.18 studied by Franx (1993) bears a strong
morphological resemblance to Coma-D216. An image of this galaxy is
shown in Franx \etal (1997), revealing it to have spiral arms surrounding
a large bulge.

\subsection{Late-Type Spiral Disks}

\noindent {\it D148 in DC2048-52:}\\
Multi-fiber spectroscopy of Dressler 148 in DC2048-52 (heferafter DC2048-D148)
revealed a PSB spectrum.  However, the S/N in the region of [OII]$\lambda$3727
is sufficiently low that we cannot exclude a small equivalent width in that
emission line, hence a small amount of ongoing star formation.  DC2048-D148
is classified as S0 on ground-based images.  The WFPC2 images (see Fig. \ref{pic3})
reveal a notably
different morphology.  A central light cusp is surrounded by a low surface
brightness spiral structure out to a radius of $\sim$7\arcsc, or $\sim$7 kpc.
Faint HII regions can be seen within a largely flocculent spiral arm pattern.
The radial luminosity profile is dominated by the central light cusp in the
inner $\sim$1\arcs in radius, where an r$^{1/4}$ law profile is evident
(see Fig. \ref{psb2_pro}).  For radii greater than $\sim$3\arcsc, an exponential disk is the
primary contributor.  The color is essentially constant in the central 1\arcsc,
and then reddens by about 0.2 mag in B--I out to a radius of $\sim$4\arcsc.
The mean B-I color is red for a late-type spiral, as is the case
for D192 below.

\noindent {\it D192 in DC2048-52:}\\
Multi-fiber spectroscopy of Dressler 192 in DC2048-52 (hereafter DC2048-D192)
reveals strong Balmer absorption lines with weak ($\sim$4--5 \AA \ equivalent width)
emission in [OII]$\lambda$3727.  We do not have a spectrum in the red that can
distinguish whether this emission is due to a small amount of ongoing star
formation or to an AGN.  DC2048-D192 was classified as PSB in Caldwell \&
Rose (1997), due to the very strong Balmer absorption and only weak 
[OII]$\lambda$3727 emission.  However, the WFPC2 images indicate that star 
formation is still ongoing in the central $\sim$1.25\arcs ($\sim$1.25 kpc)
in radius.  As can be seen in Fig. \ref{pic3} a central bright nuclear component is
surrounded by a relatively bright star forming region, which is punctuated by
HII regions and/or bright star clusters and dust lanes.  A fainter asymmetric
spiral pattern can be traced out to $\sim$3.75\arcsc, or $\sim$3.75 kpc in
radius.  The luminosity profile for DC2048-D192 (Fig. \ref{psb2_pro}) is dominated by an
exponential disk beyond $\sim$3\arcsc, while a more centrally concentrated
component is the primary contributor at smaller radii.  The radial color map
(Fig. \ref{psb2_pro}) shows the nucleus to be slightly redder than the surrounding region,
except for a red area at a radius of $\sim$2\arcs from the nucleus, at the
location of an apparent major dust lane in the image.

\section{Discussion}

At this point we can summarize our knowledge of the SB and PSB 
galaxies in nearby clusters as follows.  From ground-based multi-fiber
and long-slit spectroscopy we have found that some galaxies, which from
ground-based images are classified as early-type galaxies, have signs of
unusual star formation histories, both in terms of the star formation
being elevated in the recent past and that star formation being
centrally concentrated.  The ground-based images indicate that all of
these galaxies are disk galaxies.  From the HST images, we find that
the SB galaxies have spiral structure.  In some cases (Coma-D15) the spiral
structure is restricted to the inner couple of kpc, while in others (Coma-D45
and Coma-D100) it is global, but in all cases the star formation
is centrally concentrated.  Most
of the PSB galaxies have smooth morphologies, perhaps because their time
since the cessation of the starbursts is already $\sim$1 Gyr (Caldwell et al.
1996). However
one PSB galaxy (Coma-D61) exhibits an X-shaped bulge, which is interesting since
the starburst age of this galaxy is relatively young, at $\sim$0.5 Gyr.
None of the observed galaxies show large collections of young star 
clusters such as are seen in some merger products 
(Holtzmann \etal 1992,  Miller \etal 1997), which may be an indication that
major mergers are not the mechanism involved here.

The two late-type spiral galaxies are a cause for uncertainty.  These might
be related to anemic galaxies, in which a starburst has not actually occured;
rather, the normal disk star formation in such galaxies has been truncated.
More detailed spectroscopy than is currently available is needed to sort out 
this problem.

Although only nine SB/PSB have been studied so far, it seems reasonable
to propose an evolutionary sequence, for example, from Coma-D15 to 
Coma-D61
to Coma-D99.  Thus the starburst which is centrally located but extended
would eventually fade, resulting in a galaxy with smoother appearance, 
but with spectroscopic evidence for the 
past starburst.  We caution that precursors might well have a
large range in morphologies, and thus the resultants might also have a
range in appearances.  The central issue we now address is what these
observed morphologies allow us to conclude about the source of
the SB/PSB activity in nearby rich clusters.

The mechanisms that have been proposed to explain the rapid evolution in
galaxies taking place in distant clusters tend to fall into two basic
categories: external gravitational perturbation of some kind, (e.g.,
tidal interaction or merging) or a gas removal mechanism (e.g., ram
pressure stripping or ablation).  Ram pressure stripping is undoubtedly
an important factor in the evolution of cluster galaxies, and probably
plays a major role in the observed HI depletion of spiral disks (e.g., Gunn \& 
Gott 1972; Cowie \& Songaila 1977; Dressler 1985).
However, given the central concentration of the
star formation and the non-axisymmetric morphologies present in some
cases, which are difficult to understand in terms of a ``continuous''
gas removal process, we naturally tend to consider the external
gravitational perturbation mechanism as the more likely of the two
possibilities as a source of the observed central star formation.
Such a perturbation could arise from a single strong tidal interaction
and/or merger from either an equal mass perturber or a minor
mass galaxy. Alternatively, the tidal perturbation could be produced by
the global cluster gravitational potential (Byrd \& Valtonen 1990; Valluri 
1993; Henriksen \& Byrd 1996), or, in the case of ``galaxy
harassment'', from the combined effects of many discrete minor disturbances and
of the global cluster potential (Moore \etal 1996, 1998).

\subsection{Tidal Interaction/Mergers Between Equal Mass Galaxies}

We first consider strong tidal interactions and/or mergers of equal mass
galaxies.  Observations of galaxies in distant clusters have revealed
the signatures of major encounters/mergers in the form of long tidal tails
in a substantial number of cases (e.g., Lavery \& Henry 1988; Couch \etal 1998
and references therein).
These signatures can be taken as evidence for such mergers, although not all
the observed cases are clearly due to mergers (Couch \etal 1998).  
However, among the SB/PSB galaxies that we have imaged in nearby
clusters, there are no signs of such major interactions/mergers; we find
no cases where long tidal tails are evident, nor do we find
signs of a major companion with similar radial velocity.  In the three SB/PSB
cases where nearby ``companions'' do appear to be present, two ``companions''
have in fact very different velocities. Coma-D99 and Coma-D100 have a
radial velocity difference of 4700 km/sec and (2) the nearby
``companion'' to Coma-D15 is a background galaxy with z=0.127.
In the third case, DC2048-D216, the ``companion'' has no measured
redshift, but that galaxy itself is a normal looking late-type spiral,
with no indication of tidal interactions.  As well, DC2048-D216 has a
smooth appearance indicating that a significant time has elapsed since 
the putative tidal iteraction would have occured.
Thus we do not consider strong tidal interactions or mergers between
equal mass galaxies to be the primary mechanism for inducing starbursts
in the nearby rich clusters.

\subsection{Minor Mergers}

Mergers between a galaxy and a low-mass companion (hereafter referred to
as a minor merger) provide a more promising prospect.  Mihos \&
Hernquist (1994) and  Hernquist \& Mihos (1995) have carried out numerical 
simulations of minor mergers
which include gas dynamics and star formation.  A key result of their 
simulations is that gas in the main galaxy loses angular momentum due to
torques exerted by non-axisymmetric disturbances in the stellar component which
are generated by the tidal passage and infall of the satellite galaxy.  As a
result, a great deal of gas piles up in the central region of the main galaxy,
leading to a burst of star formation there.  The central concentration of the
star formation and the magnitude and duration of the star formation burst depend
on the bulge-to-disk ratio of the main galaxy.  However, in general the star
formation is concentrated within the central $\sim$1 kpc of the main galaxy, and the
starburst lasts for typically $\sim$100 -- 150 Myr (Mihos \& Hernquist 1994).  
Some of their simulations
bear a striking resemblance to D100 and D15 in Coma.  Specifically, in
Fig. 4 (for simulation of the gas) and Fig. 3 (for simulation of the stars)
of Hernquist \& Mihos (1995), at a time step of 43.2 the main galaxy looks
very similar to D100.  At time steps greater than 80, there is considerable resemblance
to D15.  In addition, both Whitmore \& Bell (1988) and Mihos et al.
(1995) have discussed X shapes in the bulges of galaxies and have 
proposed that they are produced by the disturbance created when a minor
merger takes place.  The Mihos et al. (1995) simulation of a minor
merger viewed edge-on (see their Fig. 3)  is a close match to our
{\it HST} 
image of D61 in Coma.  Finally, as can be seen from the edge-on views of
minor merger simulations published in Fig. 3 of Hernquist \& Mihos (1995),
minor mergers heat the disk of the main galaxy, but do not lead to its
disruption, which is in agreement with the fact that the PSB galaxies in Coma
and DC2048-52 still exhibit surviving disks.

\subsubsection{Frequency of Minor Mergers}

Given that the minor merger simulations of Mihos \& Hernquist are capable of
generating the kind of morphological signatures and central star formation 
that characterize some of our {\it HST} images, we evaluated the possibility that 
minor mergers are the {\it primary\/} source of the star forming early-type 
galaxies in nearby clusters.  Specifically, we are interested in reproducing
the situation in the SW region of the Coma cluster, where many PSB galaxies
were found by Caldwell \etal (1993), a spatial/kinematic subcluster has been
clearly identified (Colless \& Dunn 1996; Biviano \etal 1996), and a secondary 
peak in the cluster
x-ray emission has been discovered (Watt \etal 1992; Briel \etal 1992;
White \etal 1993).  While there is much current debate concerning whether this
subcluster, with its associated x-ray and starburst phenomena, is currently
falling into the main Coma cluster or has already passed through (e.g.,
Burns \etal 1994; Colless \& Dunn 1996), for the purposes of this discussion
we will assume the latter case.  Thus we speculate that the starbursts in the
Coma PSB galaxies were triggered by minor mergers which occurred when the
subcluster encountered the numerous small galaxies in the central part of the
Coma cluster.  We tested this hypothesis by running N-body
simulations of colliding clusters of galaxies using the publicly available
NEMO stellar dynamical software (Teuben 1995).  
The simulations are similar to those described in Caldwell \& Rose
(1997), in which the evolution of a subcluster falling into a main cluster
is followed.  As before, we are only concerned with the gravitational 
interactions of the particles, so the hydrodynamics of the ICM are not
considered.  A detailed description of the simulations, and of how we
identified the minor merger events, is given in the Appendix.  The results of
the simulations, which are also described in the Appendix and summarized in
Table A.1, show that the frequency of minor mergers is far too low to explain
the large fraction of SB/PSB galaxies in the SW region of Coma. 

\subsection{Galaxy Harassment}

As mentioned earlier, an alternative way to produce morphological
disturbances in cluster galaxies through external gravitational
influence is via the ``galaxy harassment'' mechanism of Moore \etal (1996, 1998).
Here the combined effects of frequent impulsive perturbations from 
small galaxies in high-speed encounters and the changing global tidal
field of the cluster produce similar disturbances to the minor merger
scenario. Given that harassment is a continual
process for any cluster galaxy, we can expect that the
morphological and star formation signatures of the harassment process
should be evident in those galaxies which still have disks containing
significant amounts of HI.  Thus galaxies in 
subclusters which are infalling for the
first time into main clusters, and hence have not yet (completely) lost
their gas supply in the hostile cluster environment, are good candidates
to produce visible reaction to the harassment process.  In fact, Fujita (1998)
has recently shown from a simple model of pressure-enhanced star formation
that galaxy harassment does indeed lead to elevated rates of star
formation.  In short,
galaxy harassment could provide a natural explanation for the appearance of
the many SB/PSB galaxies in the SW region of Coma, where the NGC 4839 
subcluster appears to be interacting with the main cluster.  Of course,
one can also expect to see isolated examples of harassed galaxies, since
any disk galaxy with a gas supply that is falling into the cluster can
in principle go through the evolutionary stages that appear to be
present among the SB and PSB galaxies in Coma and DC2048-52.

Several questions remain about the applicability of galaxy harassment
to the SB/PSB galaxies in nearby clusters.  This
process has been used to primarily describe the evolution of the small
galaxies in clusters, whereas the Coma and DC2048-52 galaxies that we
have imaged are quite substantial, typically only a magnitude fainter
than L*.  It is important to evaluate the process for these more
massive galaxies, with particular regard to the end state produced.
Specifically, our HST images and
ground-based images indicate that the PSB galaxies still have a 
prominent disk, thus observationally the disks do survive the perturbation.
But do simulated harassed disks survive?  As well, the HST images indicate
that the morphologies of the SB/PSB galaxies are quite heterogeneous,
indicating that a range in precursor morphologies are being processed
into a still considerable range in resultant morphologies.  It is
important to simulate the effect of harassment on a reasonably wide
range of input disk morphologies.  
The effect of harassment on a wide range of initial disk luminosity (mass) 
must be considered as well.  In Caldwell \& Rose (1998) we found two 
low-luminosity (B$\sim -17$) Coma galaxies in a recent PSB state.  
They are clearly S0's, as indicated by their light profiles, hence
harassment must not in general lead to the destruction of disks in smaller
galaxies if it is to explain these small galaxies in Coma.
Finally, can harassment produce the
X morphology that is so evident in Coma-D61?

\section{Conclusions}

To summarize, {\it HST} WFPC2 images of nine SB/PSB galaxies in the Coma
and DC2048-52 nearby rich clusters have revealed that the morphologies of the
galaxies are rather heterogeneous.  However, the SB galaxies tend to have
ongoing highly centralized star formation in a region typically $\sim$1-2 kpc
in radius that appears as a bright ``bulge'' on ground-based images.  In
contrast, the PSB galaxies for the most part have substantially smoother 
morphologies than their SB counterparts, but with blue colors in their
central $\sim$1-2 kpc in radius.  These findings, coupled with the ground-based
spectrosocpy reported in earlier papers, are generally consistent with a picture
in which the SB galaxies are presently experiencing enhanced star formation in
their central regions, and ultimately will fade into the the smoother
morphologies represented by the PSB galaxies, with lingering evidence for
past central starbursts.
While the sample of nine galaxies is still small for drawing general
conclusions about the mechanism that triggers the starbursts, in several
galaxies we see evidence for an external tidal perturbation as the likely cause 
of the centralized starburst.   Perhaps the ``galaxy harassment''
mechanism of Moore \etal (1996) is capable of providing such tidal
perturbations.

The fact that some galaxies in nearby clusters appear to be normal 
S0's or ellipticals on ground-based images, but with the increased spatial
resolution of {\it HST} are revealed to have centrally concentrated recent
or ongoing star formation, 
has implications for morphological studies of distant galaxies.  In
the typical z$\sim$0.4 cluster for which {\it HST} images can now provide
comparable physical resolution as ground-based images of Coma cluster galaxies,
one may now question whether some galaxies classified as early-type from
{\it HST} images actually harbor ongoing or recent starbursts in their central 
regions.

Finally, to establish a clearer picture of how a galaxy is excited into a
starburst and then fades through the PSB phase into a ``normal'' early-type
galaxy, we need to increase the sample of SB/PSB galaxies beyond the nine
observed for this paper to several dozen.  A larger sample is particularly
necessary if we are to assess how evolution through the SB and PSB phase
depends on the initial morphology of the precursor galaxy.  Such information
would establish an observational framework for testing the effects of tidal
perturbations on galaxies within the dense cluster environment.

\acknowledgements

This research was supported by Grant No. GO-06773.01-95A (cycle 6) from the
Space Telescope Science Institute to the University of North Carolina and by
NSF grant AST-9320723 to the University of North Carolina.

\appendix

\section{APPENDIX}

The numerical simulations used in \S4.2.1 to assess the frequency of
minor mergers in clusters were carried
out using the routine ``hackcode1'', which is the standard NEMO N-body
integrator based on the Barnes \& Hut (1989) hierarchial tree algorithm.
The NEMO stellar dynamical software is publicly available on the World
Wide Web.
Because we would like to compare the simulations to our observations, it
is
important to know to what physical units the NEMO units correspond.
The NEMO units are defined as follows: 1) the gravitational potential of
each
cluster is $|W|=-1/2$, 2) the kinetic energy is $K=1/4$ (therefore the
cluster
is virialized), 3) the total mass of the cluster is $M=1$ and 4) the
gravitational
constant is $G=1$.  Noting that $|W|=-GM^2/r_g$ where $r_g$ is the
gravitational
radius, we find that $r_g=2.0$, and from the kinetic energy requirements,
$<v>=1/\sqrt{2}$, where $<v>$ is the rms velocity dispersion of the
galaxies.
Observations of the x-ray gas (Fusco-Femiano \& Hughes 1994) and of the
galaxy
distribution (Kent \& Gunn 1982) 
show that the core radius of Coma is
$b\sim250$ kpc
and $b\sim340-400$ kpc respectively.  Using the fact that for a Plummer
model (which we adopt here for its mathematical convenience),
$r_g=3.25b$ and an estimated distance to the Coma cluster of 100 Mpc, we
find an
average $r_g$ on the order of 1 Mpc implying that a NEMO unit of distance
is 500 kpc.
The inferred 3-dimensional rms velocity dispersion for Coma is 
approximately 1700 km/s (Colless \& Dunn 1996; Biviano et al 1996)
implying
that a NEMO unit of velocity is 2404 km/s. 
A unit of time in NEMO is defined
as 0.4 crossing time, which, for the Coma cluster with a crossing time of
1 Gyr,
corresponds to 0.4 Gyr.

In general, the simulations consisted of two clusters: a main
cluster with $\sim$5000 particles and a less massive subcluster with 256
particles.  The point of the simulations was to assess how large a fraction of
the galaxies in the infalling subcluster would experience a minor merger with
one of the small galaxies in the main cluster.  The Mihos \& Hernquist (1994)
and Hernquist \& Mihos (1995) simulations used a mass ratio
between large galaxy and satellite galaxy of 10:1.  While it is not entirely
clear what is the minimum mass ratio that can still produce a large tidal effect
on the larger galaxy, we adopted a ratio of 20:1 as a reasonable estimate.
Consequently, our simulations are designed to look for minor mergers between
``large'' galaxies in the subcluster and ``small'' galaxies in the main
cluster.  Given that the typical PSB galaxy in the SW region of the Coma
cluster has B = 16.5, we expect the typical minor merger to be with a main
cluster galaxy with B = 19.75.  From recent analyses of the Coma cluster
luminosity function (Bernstein \etal 1995; Lobo \etal 1997; Secker \etal 1997)
we estimate there to be approximately 1200 galaxies with B $<$ 19.75.  Hence
in simulating the Coma cluster with $\sim$5000 particles we overestimate
the frequency of minor mergers by a factor of approximately 4.  However, by
using such a large number of particles in both cluster and subcluster we
obtain better statistics on the merger frequency and then can adjust the
merger rates accordingly.  

In our simulations each cluster was constructed from a virialized Plummer model
with a softening parameter for each particle of 0.05 NEMO units (equivalent to
25 kpc) to account for
the extended mass distributions of the individual galaxies.
The two structures were initially placed 4 distance units apart
with a variety of purely radial subcluster infall velocities ranging from
1/4 parabolic to 1/16 parabolic.  In most cases, the mass ratio of the two
clusters was 4:1.  The subcluster was constructed of only one type of particle
each with a mass of $3.91$x$10^{-3}$ NEMO mass units, so that the total mass of
the subcluster was 1.  For the main cluster we used 5120 particles each with
a mass of $1.95$x$10^{-4}$, i.e., approximately 20 times less massive than the
particles in the subcluster.  Then to bring the total mass of the main cluster
to four times larger than the subcluster, we inserted 100 particles each of mass
$3.0$x$10^{-2}$.
All simulations were run for 10
time units (i.e. 4 Gyr total) with 15 increments per each 0.4 Gyr time unit.
At times near 4.5, the subcluster is just passing through the center of the
main cluster and at times near 9, the subcluster
has almost halted and is about to fall back into the main cluster for the second
time.  Because we are only interested in the first passage of the subcluster,
the simulations were not continued beyond this point.  To increase the 
statistical reliability of the results, we ran each simulation four times, with
different random initial positions and velocities for the individual particles
within the two virialized Plummer models.

For each iteration, NEMO outputs the mass, the x-y-z spatial coordinates and
the x-y-z velocity coordinates of each particle.  The particle data is then
analyzed to determine if a minor merger has taken place.  The specific
conditions under which galaxy mergers occur are still somewhat uncertain.
Analytical work carried out under the impulsive approximation (e.g., Alladin
1965; Alladin \etal 1975; White 1982 and references therein) have indicated that
tidal interactions will dissipate sufficient orbital energy to effect capture
if the original encounter velocity between the two galaxies is slightly
hyperbolic or less.  More specifically, the energy dissipation from tidal
interaction scales as:
\begin{equation}
\Delta U \sim V_p^{-2}d^{-4},
\end{equation}
where $V_p$ and $d$ are the velocity and distance separation of the two galaxies
at closest approach (Spitzer 1958; Alladin \etal 1975; White 1982).  In more
general terms, White (1982) concluded that if a pair of equal mass galaxies have
 a
relative velocity and impact parameter at infinity given by $V_\infty$ and
$p$, and with relative velocity $V_p$ and distance $d$ at closest approach,
then merging will occur for $V_\infty^2 < 2<v^2>$ and $d < 2r_g$,
where $<v^2>$ and $r_g$ are the rms velocity dispersion and gravitational
radius (as defined through the virial theorem) of each galaxy.  The
encounter parameters at infinity and at closest approach are related by
$V_\infty p = V_p d$ and $V_p^2 = V_\infty^2 + 4Gm/d$, where $m$ is the mass of
each galaxy.  For the specific case of a pair of galaxies
with $<v^2>$ = 180 \kms and m = 10$^{11}M_{\sun}$, the maximum relative velocity
at infinity that can result in a merger is then $V_\infty$ = 255 \kms, and for a
closest approach of $d$ = 25 kpc, the orbital kinetic energy term $V_p^2$ is
approximately twice the gravitational potential energy term $4Gm/d$, since
$V_\infty^2 \sim 4Gm/d$.  Hence, it appears that as a general rule a merger
will occur if the oribital kinetic energy at closest approach is less than
twice the orbital potential energy.  This is clearly a very crude
approximation, given that the actual tidal encounter is dependent on the mass
distribution within the galaxies and their internal kinematics (e.g.,
elliptical versus disk galaxy) and the angle of encounter.

Unfortunately, the above assumption of an impulsive encounter
is actually invalid in the slow encounters that lead to a merger.  However,
numerical simulations (e.g., Toomre 1977; White 1982) have tended to confirm
the general results obtained from the above analytic approximations.
As a specific case, in the Hernquist \& Mihos (1995) simulations a minor merger
was achieved when the satellite galaxy was
initially placed at 6 disk scale lengths from the large galaxy,
or $\sim$20 -- 25 kpc
for a major disk galaxy like the Milky Way (e.g., Mihalas \& Binney 1981).
We therefore adopt as the first
criterion for a minor merger that the two galaxies must come within
25 kpc (or 0.05 NEMO distance units) of each other.
For our second criterion, i.e.,  that the galaxies must have a relatively slow
encounter speed, we specify that the encounter velocity at closest approach
must be low enough that the orbital kinetic energy of the main galaxy 
plus satellite
be less than twice the gravitational potential energy, i.e., the constraint
developed above from the impulsive approximation studies.
Finally, since we are interested in the frequency of minor mergers,
the third criterion is that the encounter must take place between an infalling
subcluster member and a low mass particle in the main cluster.

The general results of the N-body simulations are that typically less than
2\% of the subcluster galaxies experience a minor merger.  Given that we 
overestimated the number of small galaxies in Coma by more than a factor of 4,
we thus predict that only 0.5\% of the galaxies in a merging subcluster would
experience a minor merger.  In contrast,
$\sim$40\% of the early-type galaxies in the SW region of Coma are
either SB or PSB (Caldwell et al 1993).  Furthermore, those
mergers that do take place in our simulations are almost exclusively among
particles in
the subcluster that are stripped off during the subcluster passage and
thus
remain behind with the main cluster.  As was discussed in Caldwell \& Rose
(1997), about 60\% of the subcluster particles become bound to the main
cluster during the subcluster passage.  It is among these particles that
the
minor mergers occur in our simulations.  Moreover, roughly 85\% of
the observed
mergers take place between times 6 and 9, i.e., at late times.  What is
clearly happening is that some of the particles left behind in the main
cluster
eventually experience a minor merger, which has little to do with the
initial pasage of the subcluster through the main cluster.  This is not
what is seen in the case
of the SW region of Coma, where the SB/PSB galaxies are associated both
spatially and kinematically (in the mean) with the subcluster.  In
addition,
the age dating of the PSB galaxies indicates that they underwent
starbursts
within a fairly narrow time frame, thus implying that the starbursts were
triggered in a well-defined event, such as passage of the subcluster
through
the main cluster.

To further test these results, we changed the merger distance criterion from 25
kpc 
to 50 kpc.  While the number of minor mergers did
increase
with this change, the number only doubled to about 4\%, with
the mergers
occuring in the same location (i.e., in the main cluster center) and at
the same times.
It may seem puzzling that the number of mergers did not quadruple when the
distance
was doubled, as might be expected since the cross sectional area goes as
the square of the
distance.  However, one must consider
the energy criterion as well.  For a galaxy
to satisfy
both the distance and the energy requirements after the distance criterion
is doubled,
it must initially have a smaller velocity so that it
still meets
the energy condition at closest approach.

For completeness, we also tested the effects of changing the energy
criterion so that
the kinetic energy of the particles at closest approach must be less than
2.5 times the
gravitational potential energy, as opposed to the factor of 2 previously 
considered.  Even under these conditions,
the number of minor mergers that occured was only near 3\%.  Likewise,
there was no
significant increase in the number of minor mergers when the infall
parameters were changed.
Thus, from our N-body
simulations, it seems very unlikely that minor mergers are the main source
of the
star formation in the early type galaxies in nearby clusters.  A summary
of the results of the simulations is given in Table A.1.

\clearpage

%
%



%

\newpage

\begin{figure}
\figcaption[rose.fig1.ps]{WFPC2 F450W and ground-based images for galaxies with
vigorous ongoing star formation.  The three frames for each galaxy are the 
WFPC2 F450W images at low and high contrast (left and center panels 
respectively) followed by the ground-based image (right panel). 
The three frames for each galaxy
are the {\it HST} images at two different contrast levels followed by the
ground-based image.  Each frame covers 11.5\arcs x 11.5\arcsc. The arrow on
each of the left panels shows the direction of North in the image; the 
orthogonal line marks the direction of East. \label{pic2}}
\end{figure}

\begin{figure}
\figcaption{{\it HST} and ground-based images for PSB galaxies
in Coma and DC2048-52 which have smooth morphologies, indicating a lack of
current star formation.  As in Fig. 1, the three frames for each galaxy
are the {\it HST} images at two different contrast levels followed by the
ground-based image. Except for Coma-D216, where the frames cover 25.6\arcsc,
each frame covers 11.5\arcsc. \label{pic1}}
\end{figure}

\begin{figure}
\figcaption{WFPC2 F450W and ground-based images for galaxies with
star formation in a late-type disk.  As in Fig. 1, the three frames for each 
galaxy are the {\it HST} images at two different contrast levels followed by the
ground-based image.  \label{pic3}}
\end{figure}
\clearpage

\begin{figure}[p]
\plotone{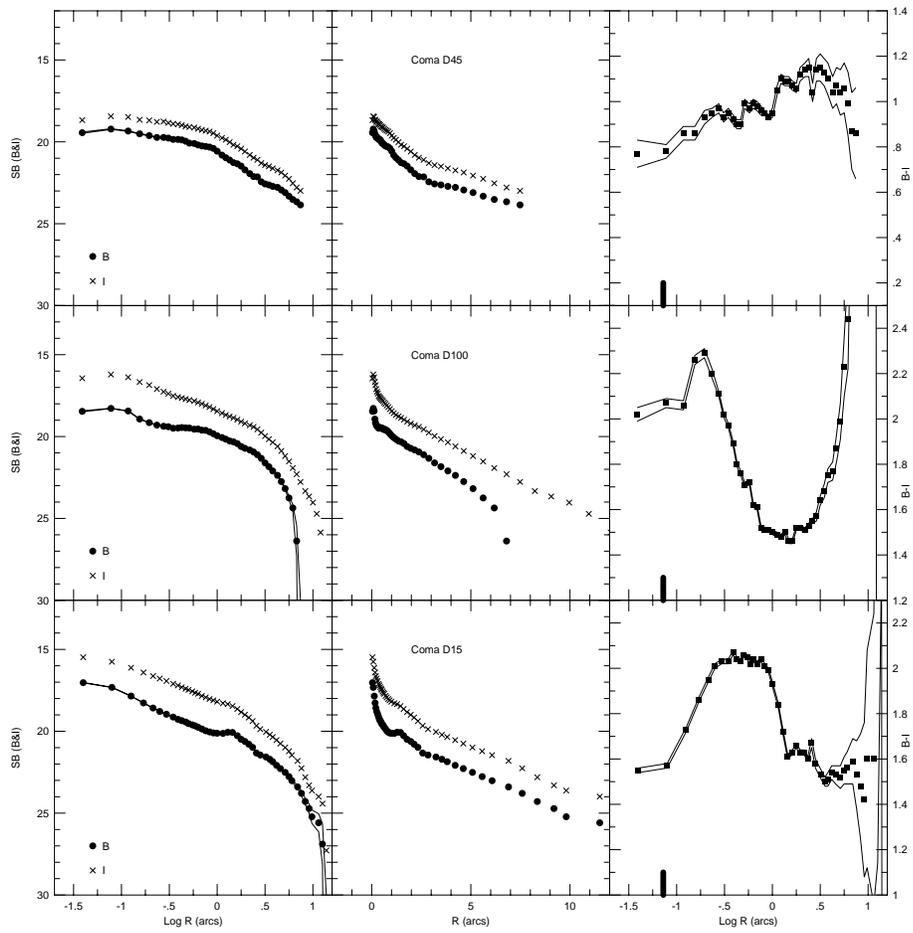}
\figcaption{Radial luminosity profiles in B (filled circles) and I (X's) for 
the SB galaxies of Fig. 1 are plotted versus
log(Radius) in left panel, and versus Radius in center panel.  The right
panel shows the B-I radial color profile plotted versus log(R).  In all panels
the thin solid lines show the 1 $\sigma$ mag errors for the B and I light profiles,
and for the B-I color profiles.  The heavy short lines at the bottom of each 
color profile shows the FWHM of stars in the frames. \label{sb_pro}}
\end{figure}
\clearpage

\begin{figure}[p]
\plotone{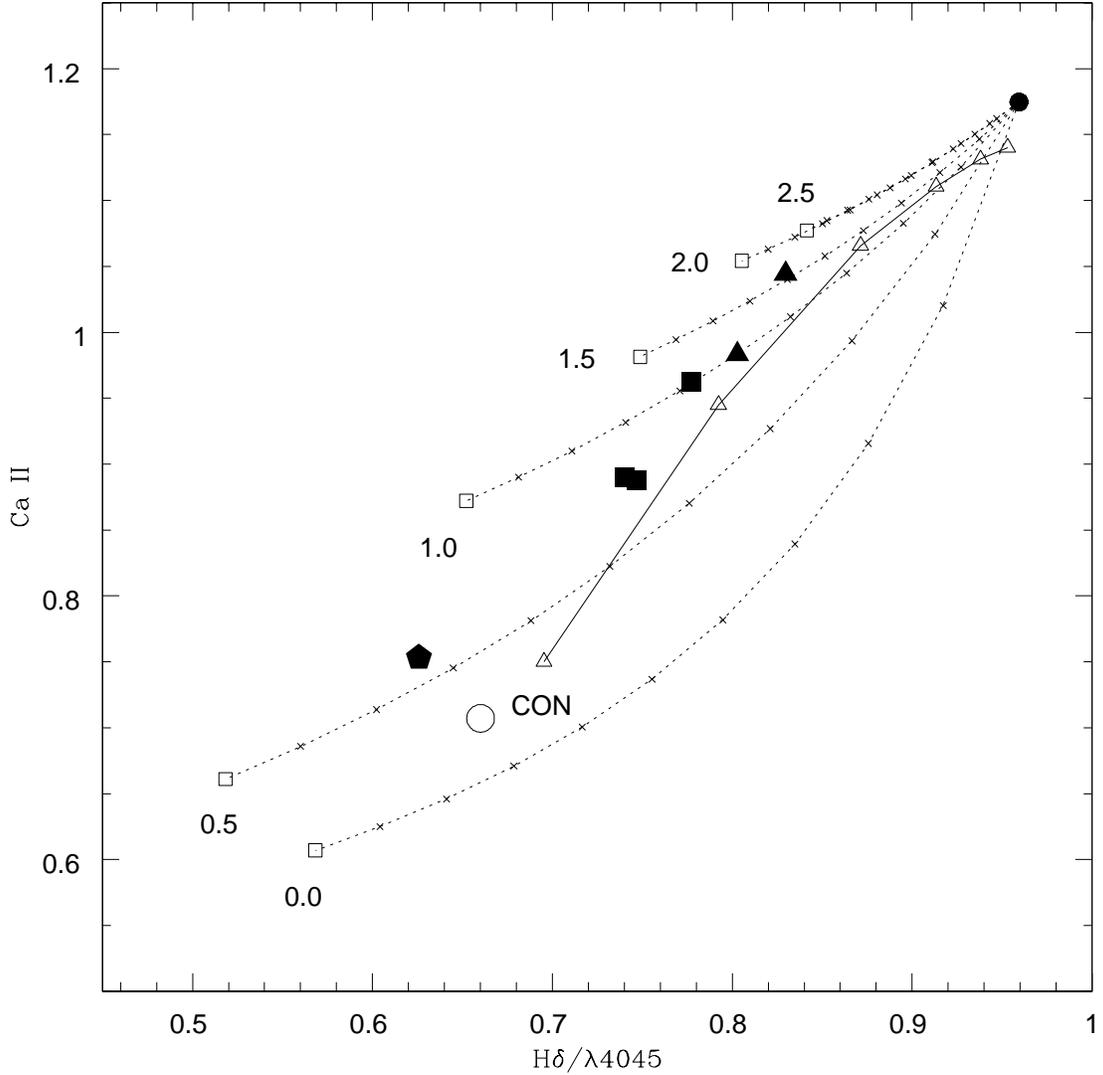}
\figcaption{Ca II vs. H$\delta$/$\lambda$4045 diagram for burst models and for 
various Coma
galaxies.  Curved, dashed lines are derived from linear combinations of
post-starburst model spectra with the observed composite spectrum of
an old galaxy population.
Unfilled squares represent Bruzual and Charlot (1995) model spectra
for a pure 0.3 Gyr long starburst that is seen at the noted times after
termination of the burst.  The small crosses designate 10\% increments in the
balance of burst versus old population light, normalized at 4000 \AA.  The
solid line marked ``CON'' represents the evolutionary track of a truncated
constant (over 15 Gyr) star formation population seen at different times after
termination of star formation.  The unfilled triangle at the bottom of the line
represents the index values right after the truncation of star formation, while
each successive triangle denotes a time step of 0.5 Gyr. The
extranuclear PSB region of Coma-D100 is plotted as an unfilled circle, 
while the PSB 
galaxy Coma-D61 is plotted as a filled pentagon. 
Also plotted are the two faint Coma galaxies GMP2903 and GMP5284 (filled 
triangles) previously studied in Caldwell \& Rose (1998), and three
PSB galaxies (filled squares) studied in Caldwell et al. (1996). \label{age_ps}}
\end{figure}
\clearpage

\begin{figure}[p]
\plotone{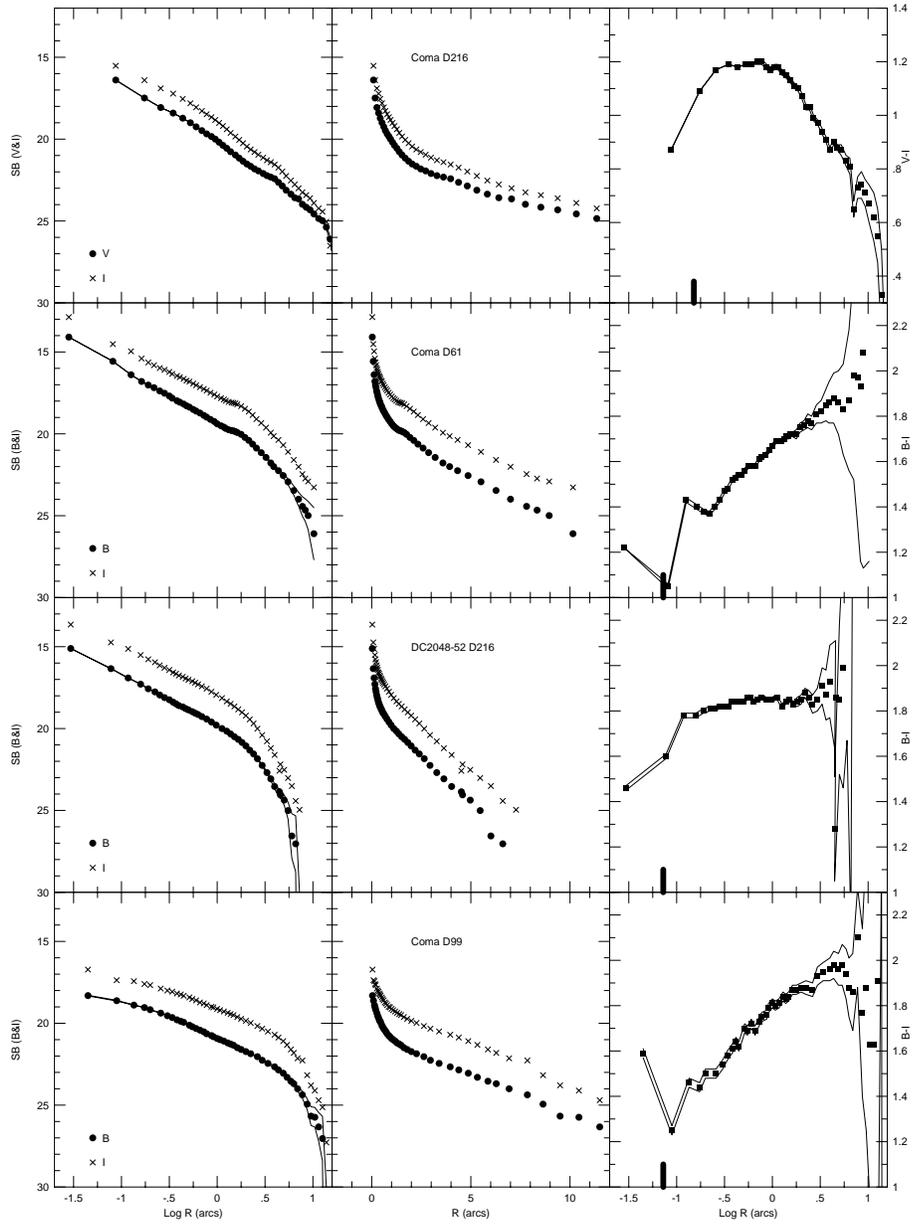}
\figcaption{Radial luminosity and color profiles in B and I for the PSB galaxies
of Fig. 2.  The symbols and panel arrangements are the same as for Fig. 4,
except that the data are for V and I (rather than B and I) in the case of
Coma-D216. \label{psb_pro}}
\end{figure}
\clearpage

\begin{figure}[p]
\plotone{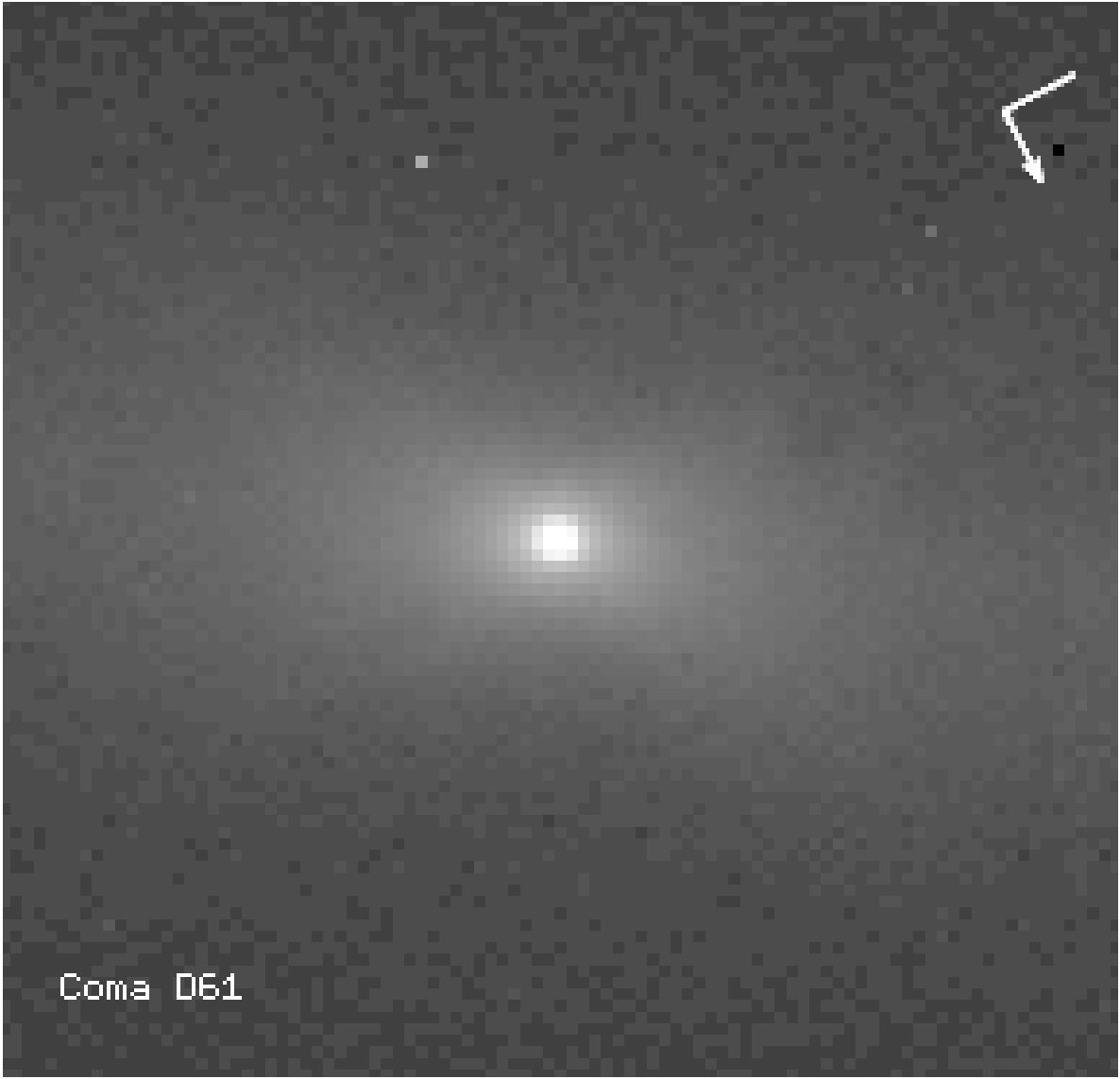}
\figcaption{Unresolved nucleus of Coma-D61 in B.  This image, 4.2\arcs across,
was created by using the Lucy-Richardson deconvolution algorithm.  It is
displayed using a non-linear lookup table. \label{d61.nuc}}
\end{figure}
\clearpage

\begin{figure}[p]
\plotone{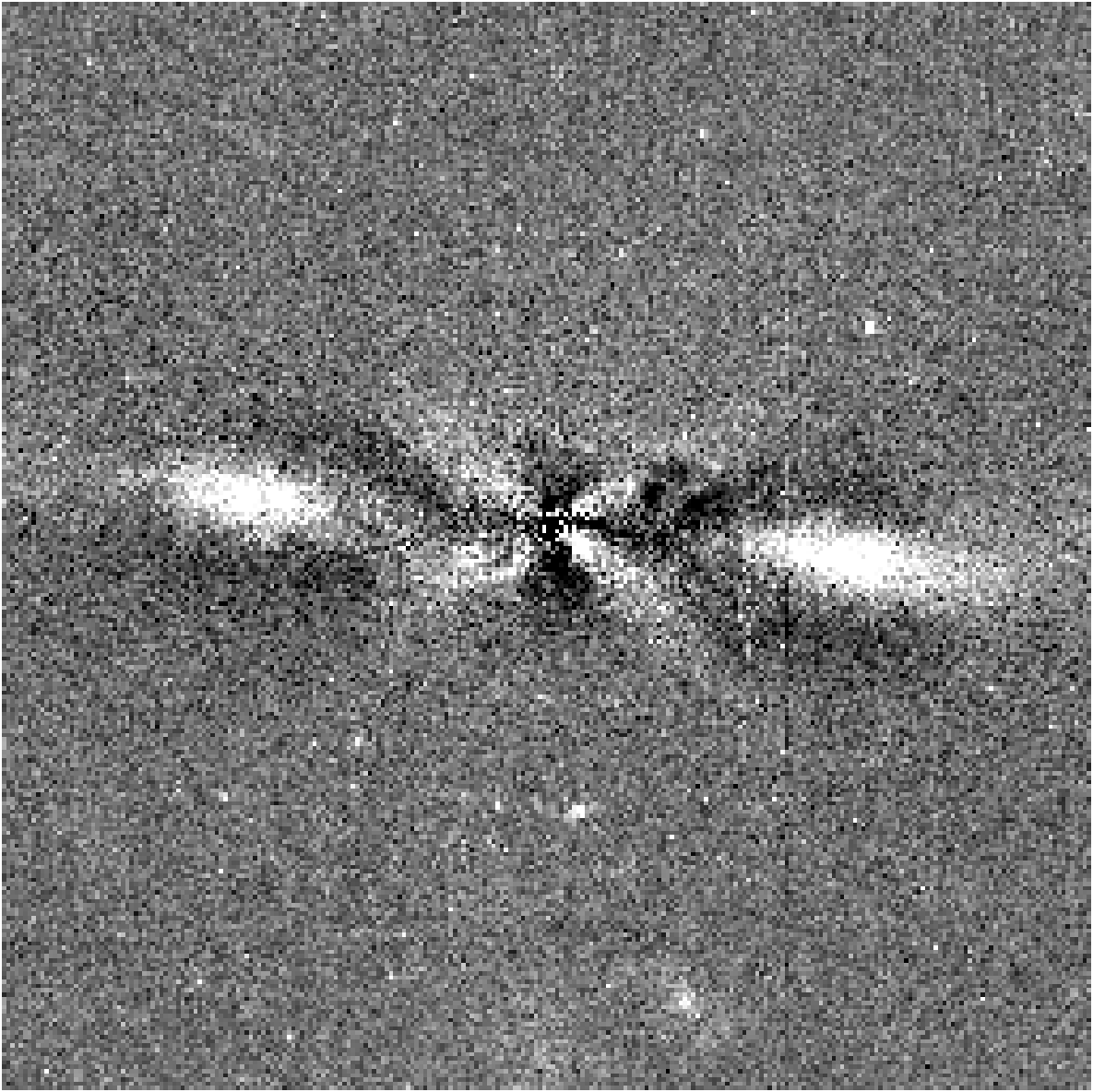}
\figcaption{WFPC2 B image of Coma-D61 with the axisymmetric bulge component
subtracted out by making elliptical isophotal fits to the bulge.  The X-shape
of the bulge is now clearly evident. \label{d61}}
\end{figure}
\clearpage

\begin{figure}[p]
\plotone{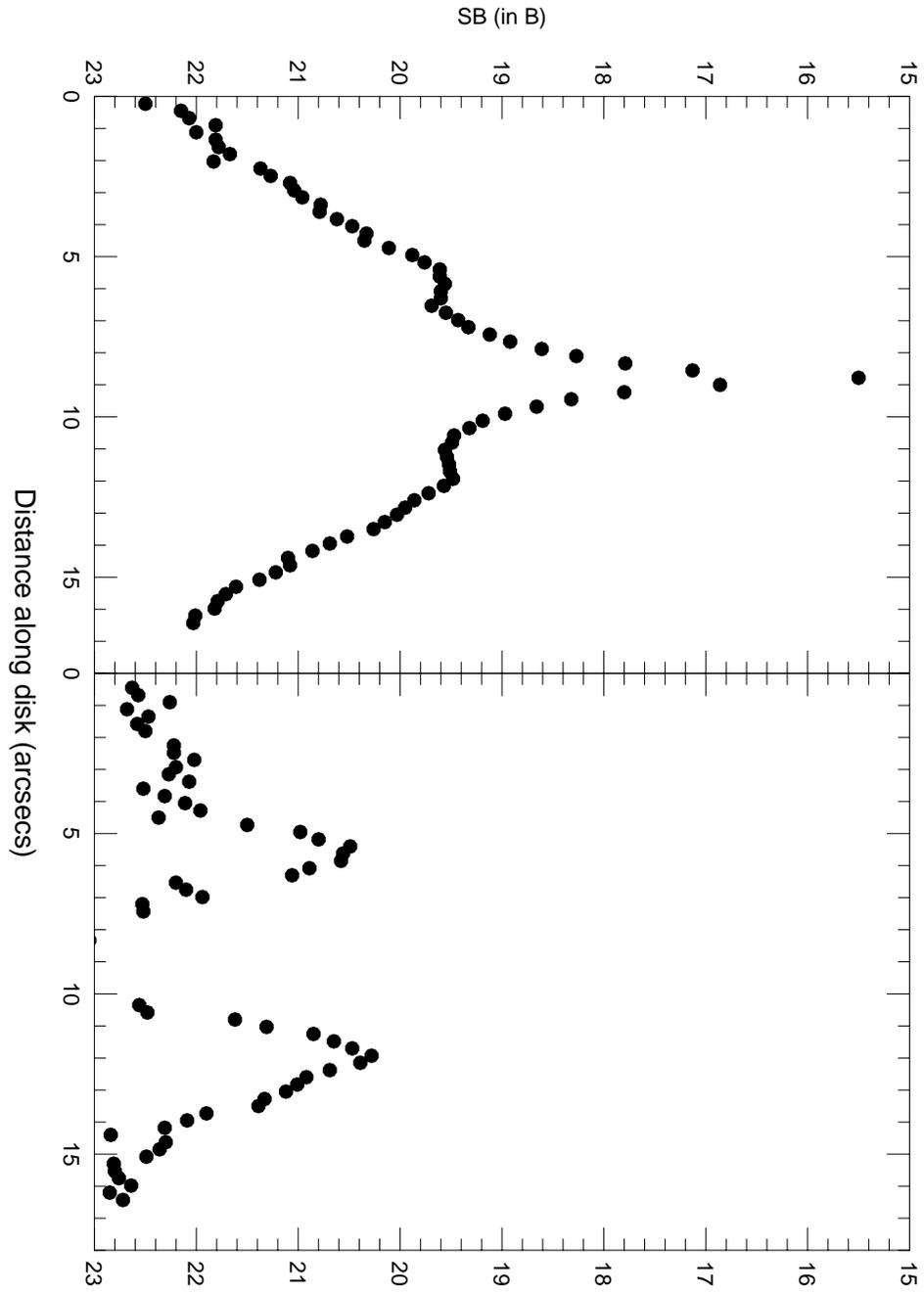}
\figcaption{Radial surface brightness profile along the major axis of Coma-D61.
The left hand panel shows the profile for the unprocessed B image, while
the right hand panel shows the remnant profile obtained after subtracting 
off the elliptical isophote fits. \label{d61_pro}}
\end{figure}
\clearpage

\begin{figure}[p]
\plotone{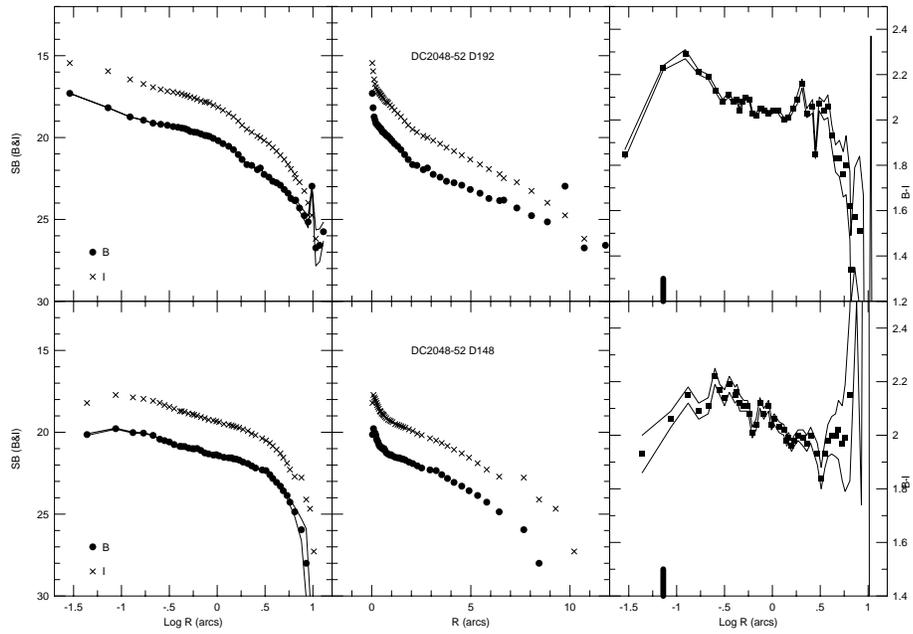}
\figcaption{Radial luminosity and color profiles in B and I for the PSB galaxies
of Fig. 3.  The symbols and panel arrangements are the same as for Fig. 4.
\label{psb2_pro}}
\end{figure}
\clearpage

\begin{figure}
\plotone{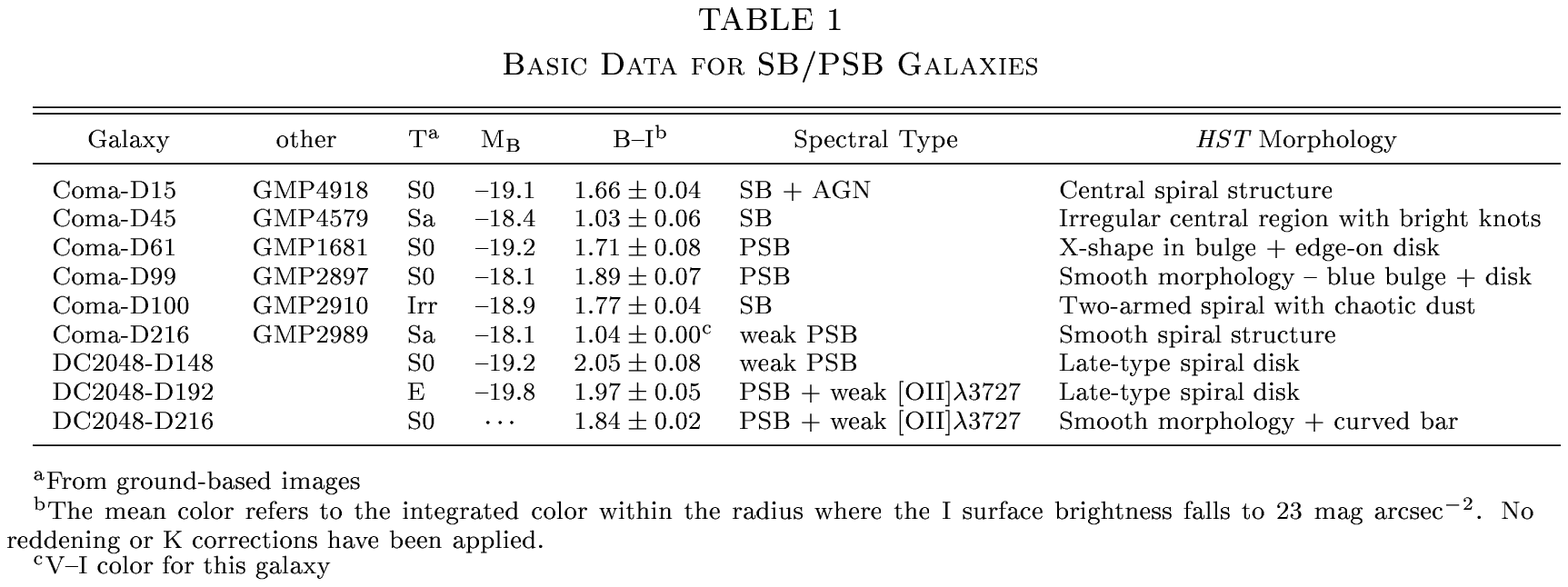}
\end{figure}
\clearpage
 
\begin{table}
\dummytable\label{tab1}
\end{table}

\begin{deluxetable}{lcccccc}
\tablenum{A.1}
\tablecolumns{7}
\tablewidth{0pc}
\tablecaption{Summary of Simulation Results}
\tablehead{
\colhead{$D_{\min}$} & \colhead{T/U} & \colhead{No. Minor} &
\colhead{\% Minor} & \colhead{\% Trapped in} & \colhead{\% at} \\
\colhead{} & \colhead{} & \colhead{Mergers} & \colhead{Mergers} &
\colhead{Main Cluster} & \colhead{t$>$6.0}
}
\startdata
   25 kpc & 2.0 & 5.3 & 2.0 \% & 85\% & 88\% \\
   50 kpc & 2.0 & 10.3 & 4.0\% & 75\% & 90\% \\
   25 kpc & 2.5 & 8.4 & 3.3\% & 85\% & 81\% \\
\enddata
\end{deluxetable}
%
%


\begin{references}
\reference{al65} Alladin, S. M. 1965, \apj, 141, 768
\reference{aps75} Alladin, S. M., Potdar, A., \& Sastry, K. S. 1975, in
Dynamics of Stellar Systems, edited by A. Hayli, (Reidel, Dordrecht), 167
\reference{and96} Andreon, S., Davoust, E., Michard, R., Nieto, J.-L.,
\& Poulain, P. 1996, \aaps, 116, 429
\reference{bpt} Baldwin, J. A., Phillips, M. M., \& Terlevich, R. 1981, \pasp,
93, 5
\reference{ba97} Balogh, M. L., Morris, S. L., Yee, H. K. C., Carlberg, 
R. G., \& Ellingson, R. 1997, \apjl, 488, L75
\reference{ba96} Barger, A. J., Arag\'on-Salamanca, A., Ellis, R. S., Couch, W. 
J., Smail, I., \& Sharples, R. M. 1996, \mnras, 279, 1
\reference{bh89} Barnes, J. E., \& Hut, P. 1989, \apjs, 70, 389
\reference{bel95} Belloni, P., Bruzual, G., Thimm, G. -J., \& Roser, H.-J. 1995,
 \aap, 297, 61
\reference{bnt95} Bernstein, G. M., Nichol, R. C., Tyson, J. A., Ulmer, M. P.,
 \& Wittman, D. 1995, \aj, 110, 1507
\reference{B96} Biviano, A., Durret, F., Gerbal, D., Le Fevre, O., Lobo, C., 
Mazure, A., \& Slezak, E. 1996, A\&A, 311, 95
\reference{bow91} Bower, R. G. 1991, \mnras, 248, 332
\reference{ble92} Bower, R. G., Lucey, J. R., \& Ellis, R. S. 1992, \mnras, 
254, 601
\reference{bhb92} Briel, , U. G., Henry, J. P., \& Bohringer, H. 1992, \aap, 259
, L31
\reference{bru98} Bruzual A. G., \& Charlot, S. 1995, \apj, in preparation
\reference{bu94a} Burns, J. O., Roettiger, K., Ledlow, M., \& Klypin, A. 1994,
\apjl,
427, L87
\reference{bo78} Butcher, H. \& Oemler, A. 1978, \apj, 219, 18
\reference{bo84} Butcher, H. \& Oemler, A. 1984, \apj, 285, 426
\reference{bv90} Byrd, G., \& Valtonen, M. 1990, \apj, 350, 89
\reference{cp89} Caldwell, N., \& Phillips, M. M. 1989, \apj, 338, 789
\reference{cal97} Caldwell, N., \& Rose, J. A. 1997, \aj, 113, 492
\reference{cal98} Caldwell, N., \& Rose, J. A. 1998, \aj, 115, 1423
\reference{cal96} Caldwell, N., Rose, J. A., Franx, M., \& Leonardi, A. 1996, 
\aj, 111, 78
\reference{cal93} Caldwell, N., Rose, J. A., Sharples, R. M., Ellis, R. S., \&
Bower, R. G. 1993, \aj, 106, 473
\reference{carl98} Carlson, M. N., et al. 1998, \aj, 115, 1778
\reference{cd96} Colless, M. \& Dunn, A. 1996, \apj, 458, 435
\reference{cb98} Couch, W. J., Barger, A. J., Smail, I., Ellis, R. S., \& 
Sharples, R. M. 1998, \apj, 497, 188
\reference{cou94} Couch, W., Ellis, R.S., Sharples, R.M., \& Smail, I. 1994,
\apj, 430, 121
\reference{cn84} Couch, W. J. \& Newell, E. B. 1984, \apjs, 56, 143
\reference{cs87} Couch, W. \& Sharples, R.M. 1987, \mnras, 229, 423
\reference{cs77} Cowie, L. L., \& Songaila, A. 1977, Nature, 266, 501
\reference{dr80} Dressler, A. 1980, \apjs, 42, 565
\reference{dr85} Dressler, A. 1985, \araa, 22, 185
\reference{dg82} Dressler, A. \& Gunn, J. E. 1982, \apj, 263, 533
\reference{dg83} Dressler, A. \& Gunn, J. E. 1983, \apj, 270, 7
\reference{dg92} Dressler, A. \& Gunn, J. E. 1992, \apjs, 78, 1
\reference{dr94a} Dressler, A., Oemler, A., Butcher, H. R., \& Gunn, J. E.
1994a, \apj, 430, 107
\reference{dr94b} Dressler, A., Oemler, A., Sparks, W. B., \& Lucas, R. A. 
1994b, \apjl, 435, L23
\reference{fab9x} Fabricant, D., Cheilets, P., Caldwell, N., \& Geary, J. 
1998, \pasp, 110, 79
\reference{fr93} Franx, M. 1993, \apjl, 407, L5
\reference{fr97} Franx, M., Kelson, D., van Dokkum, P. Illingworth, G., \&
Fabricant, D. 1997, in ASP Conference Series ``The Second Stromlo Symposium: The
Nature of Elliptical Galaxies'', eds. M. Arnaboldi, G. S. da Costa, \& P. Saha, 
116, p. 512
\reference{fuj98} Fujita, Y. 1998, astro-ph/9807120
\reference{ffh94} Fusco-Femiano, R., \& Hughes, J. P. 1994, \apj, 429, 545
\reference{ga89} Gavazzi, G. 1989, \apj, 346, 59
\reference{ga95} Gavazzi, G., Randone, I., \& Branchini, E. 1995, \apj, 438, 590
\reference{gh85} Giovanelli, R., \& Haynes, M. P. 1985, \apj, 292, 404
\reference{gg72} Gunn, J. E., \& Gott, J. R. 1972, \apj, 176, 1
\reference{hgc84} Haynes, M. P., Giovanelli, R., \& Chincarini, G. L. 1984,
\araa, 22, 445 
\reference{hb96} Henriksen, M. J., \& Byrd, G. 1996, \apj, 459, 82
\reference{hl87} Henry, J. P., \& Lavery, R. J. 1987, \apj, 323, 473
\reference{hm95} Hernquist, L. \& Mihos, J. C. 1995, \apj, 448, 41
\reference{ho92} Holtzman, J. A., Faber, S. M., Shaya, E. J.,
Lauer, T. R., Grothe, J., Hunter, D. A., Baum, W. A., Ewald, S. P., 
Hester, J. F., Light, R. M., Lynds, C. R., O'Neil, E. J., jr., \& 
Westphal, J. A. 1992, \aj, 103, 691
\reference{kau95} Kauffmann, G. 1995, \mnras, 274, 153
\reference{kent82} Kent, S. M., \& Gunn, J. E. 1982, \aj, 87, 945
\reference{kk98} Koopmann, R. A., \& Kenney, J. D. P. 1998, \apjl, 497, L75
\reference{ko82} Kormendy, J. 1982, in Morphology and Dynamics of Galaxies,
edited by L. Martinet \& M. Mayor, (Geneva Observatory), 113
\reference{lau95} Lauer, T. R., Ajhar, E. A., Byun, Y.-I., Dressler, A.,
Faber, S. M., Grillmair, C., Kormendy, J., Richstone, D., \& Tremaine, S.
1995, \aj, 110, 2622
\reference{lh86} Lavery, R. J., \& Henry, J. P. 1986, \apjl, 304, L5
\reference{lh88} Lavery, R. J., \& Henry, J. P. 1988, \apj 330, 596
\reference{lpm92} Lavery, R. J., Pierce, M. J., \& McClure, R. D. 1992, \aj,
104, 2067
\reference{leo96} Leonardi, A. J., \& Rose, J. A. 1996, \aj, 111, 182
\reference{lob97} Lobo, C., Biviano, A., Durret, F., Gerbal, D., Le Fevre, O.,
Mazure, A., \& Slezak, E. 1997, \aap, 317, 385
\reference{lucy74} Lucy, L. B. 1974, \aj, 79, 745 
\reference{mec88} MacLaren, I., Ellis, R. S., \& Couch, W. J. 1988, \mnras, 230,
 249
\reference{mb} Mihalas, D., \& Binney, J. 1981, in Galactic Astronomy: Structure
and Kinematics, Second Edition, (W. H. Freeman and Co., San Francisco), 326
\reference{mh94} Mihos, J. C., \& Hernquist, L. 1994, \apjl, 425, L13
\reference{mwh95} Mihos, J. C., Walker, I. R., Hernquist, L., de Oliveira, C.
M., \& Bolte, M. 1995, \apjl, 447, L87
\reference{mi97} Miller, B. W.,  Whitmore, B. C., Schweizer, F., \&
Fall, S. M. 1997, \aj, 114, 2381
\reference{mo96} Moore, B., Katz, N., Lake, G., Dressler, A., \& Oemler, A.
1996, Nature, 379, 613
\reference{mlk98} Moore, B., Lake, G., \& Katz, N. 1998, \apj, 495, 139
\reference{mor98} Morris, S. L., Hutchings, J. B., Carlberg, R. G., Yee, H. K.
C., Ellingson, E., Balogh, M. L., Abraham, R. G., \& Smecker-Hane, T. A. \apj,
In Press, astro-ph/9805216
\reference{mw93} Moss, C., \& Whittle, M. 1993, \apjl, 407, L17
\reference{mw95} Moss, C., Whittle, M., Pesce, J. E., \& Socas-Navarro, H.
1995, Astro Lett. and Communications, 31, 215
\reference{pvdk91} Pickles, A. J. \& van der Kruit, P. C. 1991, \aaps, 91, 1
\reference{rs95} Rakos, K. D. \& Schombert, J. M. 1995, \apj, 439, 47
\reference{rich72} Richardson, W. H. 1972, JOSA, 62, 52
\reference{shp97} Secker, J., Harris, W. E., \& Plummer, J. D. 1997, \pasp, 109,
1377
\reference{sou88} Soucail, G., Mellier, Y., Fort, B., \& Cailloux, M. 1988, 
\aaps, 73, 471
\reference{sp58} Spitzer, L. 1958, \apj, 127, 17
\reference{te95} Teuben, P. J. 1995, in Astronomical Data Analysis Software and
Systems IV, edited by R. Shaw, H. E. Payne, \& J. J. E. Hayes, PASP
Conf. Ser. No. 77 (ASP, San Francisco), 398
\reference{th88} Thompson, L. 1988, \apj, 324, 112
\reference{to77} Toomre, A. 1977, in Evolution of Galaxies and Stellar 
Populations, edited by R. B. Larson \& B. M. Tinsley, (Yale University
Observatory, New Haven), 401
\reference{val93} Valluri, M. 1993, \apj, 408, 57
\reference{vei87} Veilleux, S. \& Osterbrock, D. E. 1987, \apjs, 63, 295
\reference{wat92} Watt, M. P., Ponman, T. J., Bertram, D. et al. 1992, \mnras, 2
58, 738
\reference{wh82} White, S. D. M. 1982, in Morphology and Dynamics of Galaxies,
edited by L. Martinet \& M. Mayor, (Geneva Observatory), 291
\reference{wbh93} White, S. D. M., Briel, U. G., \& Henry, J. P. 1993, \mnras, 2
61, L8
\reference{wb88} Whitmore, B. C., \& Bell, M. 1988, \apj, 324, 741
\reference{whit93} Whitmore, B. C., Schweizer, F., Leitherer, C., 
Borne, K, \& Robert, C. 1993, \aj, 106, 1354
\reference{whit97} Whitmore, B. C., Miller, B. W., Schweizer, F., \&
Fall, S. M. 1997, \aj, 114, 1797
\reference{wi94} Wirth, G. D., Koo, D. C., \& Kron, R. G. 1994, \apjl, 435, L105

\end{references}
\end{document}